# Wireless strain and temperature monitoring in reinforced concrete using Surface Acoustic Wave (SAW) sensors

Pierre Jeltiri[1,2] *, Firas Al-Mahmoud[1], Rémi Boissière[1], Baptiste Paulmier[1], Tony Makdissy[1], Omar Elmazria[1] **, Pascal Nicolay[2]**, and Sami Hage-Ali[1] ***

[1] Université de Lorraine, CNRS, IJL UMR 7198, F-54000 Nancy, France
[2] Carinthia Institute for Smart Materials (CiSMAT), Carinthia University of Applied Sciences, Villach, Austria
* IEEE Student Member, ** IEEE Senior Member, *** IEEE Member,

**Abstract—** Monitoring the health of civil engineering structures using implanted deformation, temperature and corrosion sensors would further improve maintenance and extend the service life of those structures. However, sensor integration poses a number of problems, due to the presence of cables and on-board electronics. Passive, wireless SAW sensors offer a very promising solution, here. We used commercial SAW devices mounted on steel rebars to carry out an initial feasibility study. Without cables or embedded electronics, we were able to measure the deformation of a concrete beam subjected to bending load. We were also able to measure the temperature continuously over a three-week period.

**Index Terms—** Strain and temperature monitoring, Surface Acoustic Wave (SAW), wireless sensors

## I. INTRODUCTION

Structural Health Monitoring (SHM) of concrete structures is a challenge for the future, with an ever-growing need for safer and more resilient buildings, bridges and tunnels. In terms of early damage detection, measuring the strain is key. Temperature and corrosion [1] are also important parameters, for the SHM of concrete structures.

Several solutions already exist to measure the surface deformation state of civil engineering structures. This includes piezoresistive strain gauges, vibrating wire strain gauges [2] and various types of fiber-optic strain sensors [3, 4]. The first two are mature solutions, developed already in the first half of the 20$^{th}$ century. These are rugged and affordable sensors. Fiber-optic sensors are a more recent but very promising solution. This family of sensors is divided in several groups, depending on their working principle (e.g., extrinsic Fabry-Perot interferometry, fiber Bragg-Grating…). A single fiber mounted along the structure can be used to measure deformation at dozens of different points. However, fiber-optic strain sensors are expensive and challenging to implement, due to their fragility. If the fiber breaks, all the measurement points are also lost, with no hope of replacement. All these different solutions share one common drawback: they are either wired or their wireless implementation is not possible without embedded electronics and power supply. This prohibits long-term use of the sensors, in concrete. Solutions based on wirelessly interrogated resistive strain gauges are for instance limited by the energy consumption of the device [5].

In this context, Surface Acoustic Wave (SAW) sensors constitute an attractive solution: they are compact (with a footprint of less than 1cm$^2$, which makes it possible to measure strain at very well-defined locations), fully passive, batteryless and can be queried wirelessly. They have already been implemented as wireless temperature and strain sensors in numerous industrial settings [6-8].

SAW sensors are based on piezoelectric materials (e.g., quartz, langasite, lithium niobate, lithium tantalate) on top of which interdigital transducers with a spatial period λ are deposited, λ being the wavelength of the acoustic wave. The operating frequency (f) of the SAW devices is given by f=c/λ, where c is the velocity of the acoustic wave. Under mechanical strain or when temperature changes, both the spatial period λ and the elastic parameters of the propagating medium (hence the velocity c) vary, causing a frequency change. This frequency can then be tracked remotely using a commercial interrogation unit or a vector network analyzer (VNA), operating for instance in RF ISM bands (see Fig. 1). However, since SAW devices are passive, the interrogation distance is relatively short (of the order of a few meters). Their level of performance is also degraded by anything that interferes with the transmission of an RF signal. These are two of the main limitations of the SAW sensors technology.

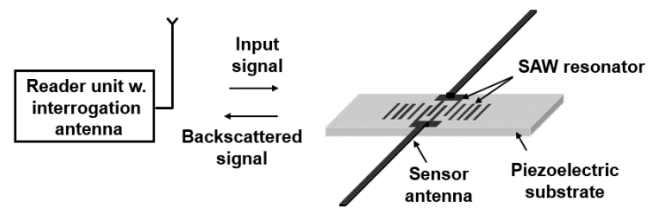

Fig. 1. Wireless interrogation principle of a SAW resonator.

The goal of this paper is to provide a first technical assessment of the potential of SAW sensors as embedded sensors, for Structural Health Monitoring Applications. To this end, wireless SAW temperature and strain sensors, as well as wired reference sensors, were embedded into a reinforced concrete beam and extensively tested in various conditions. In section II, the design of the beam and the implementation procedure of the sensors are described. In section III, we focus on the wireless interrogation of one embedded SAW temperature sensor. In section IV, we describe the results of our bending strain measurements, and discuss some of the observed issues.

Corresponding author: Pierre Jeltiri (pierre.jeltiri@univ-lorraine.fr).

## II. INSTRUMENTED REINFORCED CONCRETE BEAM: MATERIALS AND METHODS

### A. Design of the beam and sensor implementation

First, a Reinforced Concrete (RC) beam was designed. To keep the weight under 80kg, its dimensions were chosen to be relatively small, with a length of 1.2 m and a rectangular cross-section of 155 mm x 200 mm. The reinforcements consisted in two 10 mm-diameter ribbed bars in the tensile zone, and two 8 mm-diameter ribbed bars in the compression zone. 8 mm-diameter ribbed steel bars 100 mm apart were used, for the stirrups (see Fig. 2a).

The two 10 mm-diameter tension reinforcements were equipped with various temperature and strain sensors, mounted on flattened sections of the rebars (see Fig. 2b). Rebar 1 was instrumented with an unpackaged 2 mm x 1 mm x 350 µm 869 MHz SAW resonator from SAW Components Dresden (SCD), serving as a strain sensor and connected to a Taoglas ceramic patch antenna, as shown in Fig. 2b. As a reference, a 350 Ω piezoresistive strain gauge from Micro-Measurements was used. Both sensors were glued on the rebar using a cyanoacrylate strain gauge adhesive (M-Bond 200).

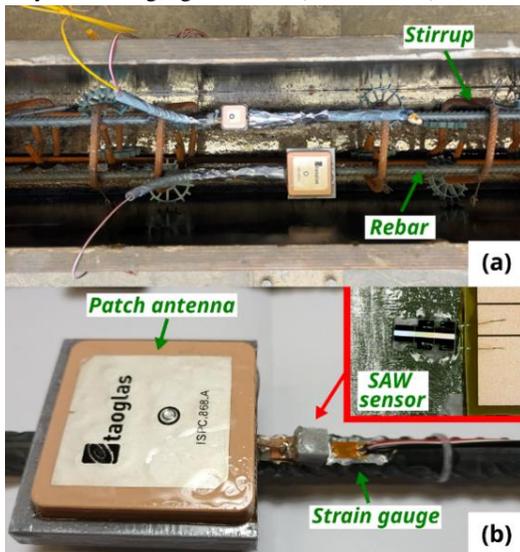

Fig. 2. (a) Instrumented beam before concrete casting (upside down view) (b) Instrumented rebar 1. Inset: SAW device glued on metal flat.

On rebar 2, we mounted:
- a wired SAW strain sensor identical to the one mounted on rebar 1, connected to the read-out electronics using a RF coaxial cable,
- a wireless, commercial SCD SAW temperature sensor at 2.459 GHz, connected to a Taoglas 2.45 GHz ceramic patch antenna. The temperature sensor was still in its original ceramic package and was only loosely bonded to the rebar. It was therefore mechanically decoupled from the rebar,
- two reference sensors: a thermocouple and another piezoresistive strain gauge.

### B. Sensor protection and beam preparation

Self-Compacting Concrete (SCC) was chosen. Since it does not require vibration/shaking after casting, the use of SCC results in a "gentler" overall fabrication procedure. Still, the sensors need to be protected against highly chemically reactive liquid concrete (pH=13), and against direct shocks with aggregates and stones. To this end, both bare-die wired and wireless strain sensors were protected with 3D-printed housings. All parts in contact with concrete were then covered with epoxy glue. It was found in earlier work that this material provides good short-term protection against corrosive effects from concrete. Finally, to prevent mechanical damage during casting, the instrumented zone was wrapped in butyl rubber and aluminum duct tape. The antennas were only covered with a 1 mm-thick epoxy layer. The commercial 868 MHz and 2.45 GHz antennas were directly used, without any modification. The expected downwards detuning due to the presence of concrete around them was neglected. Note that it should be possible to fine-tune the antennas to compensate for this detuning, considering the dielectric properties of dry concrete ($\varepsilon_r$=4.7). The concrete is rather lossy (RF tan δ =0.13) and this resistive loss will impact the RF link budget in proportion to concrete thickness. The instrumented tension reinforcements are shown in figure 2a. A wooden mould with the correct dimensions was used, to cast the RC beam. The compressive strength (fc') of the used SCC was determined by laboratory testing of three hardened concrete cylinder specimens, which were fabricated at the same time as the beam. The cylinders have a diameter of 160 mm and a height of 320 mm. They were tested after 21 days, in compression. They showed an average fc' of 19.56 MPa.

### C. Measurement electronics and wireless interrogation

The $S_{11}$(dB) parameters of the wired and wireless SAW strain sensors at 869 MHz were measured using a Rohde & Schwartz ZNLE6 VNA. The wired sensor was connected to the VNA via coaxial cable. To read the wireless sensor, a 868 MHz commercial patch antenna was used, instead. Here, the operating frequency of the sensors is the local minimum of this interrogation antenna $S_{11}$. To read the wireless SAW temperature sensor, we used a commercial reader from SCD, connected to a commercial patch antenna at 2.45 GHz. The SCD reader provides the SAW frequency, the temperature, the amplitude of the detected SAW peak in the frequency domain and the Signal-to-Noise ratio (SNR). To read the strain gauges and the thermocouple, we used respectively a model D4 conditioner from Micro-measurements and a multimeter. A schematic diagram of the experimental setup is presented in Fig. 3.

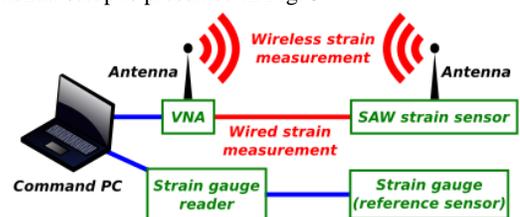

Fig. 3. Schematic diagram of the experimental setup used to measure strain inside a concrete beam, using a 869 MHz SAW sensor.

## III. WIRELESS TEMPERATURE MONITORING

It was possible to acquire the temperature, at a 2.5 cm interrogation distance, right after concrete casting (even though concrete was still liquid, with high dielectric constant and very high conductive loss). We started recording the data from the thermocouple and SAW temperature sensor after three hours, and over 21 days. The mould was removed on day 9, which required to temporarily stop the recordings.

For unknown reasons, the thermocouple data became incoherent after 10 days. The (wireless) SAW temperature measurements are shown in figure 4a. It was possible to read the SAW continuously, within fresh and hardened concrete. The observed precision was close to ±0.5°C in fresh concrete, and ±0.1°C in hardened concrete. The significant temperature increase observed within the first 24 hours was due to exothermic reactions that occur inside concrete, in the early hydration stage. The subsequent temperature increase was due to a warming weather over the measurement period, with day-night oscillations. The noise-related measurement error also decreased with time. This can be explained by the decrease in water content, during concrete hardening. This trend was also confirmed by SNR and peak amplitude measurements. Using the commercial SAW temperature sensor, we were also able to verify the possibility of reading integrated SAW sensors at a distance of several tens of centimeters, above the concrete surface. The set-up used for these tests is shown in figure 4b. During the tests, we were able to achieve a maximum reading distance of around 1 m, which is an interesting and promising result for future applications of SAW technology in the SHM field.

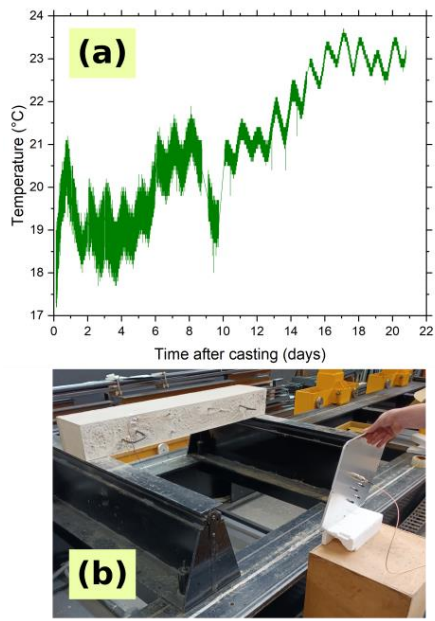

Fig. 4. (a) Temperature measurement, using the integrated commercial SAW sensor, operating at 2.45 GHz. (b) Wireless reading setup of the commercial SAW sensor, operating at 2.45 GHz.

## IV. WIRED AND WIRELESS STRAIN SENSING

### A. Manual bending tests with metal weights

We could start the tests after one week. The wired SAW strain sensor on rebar 2 was characterized, first. A weight of up to 200kg was placed on top at the mid-span of the beam, in 40kg increments. This generated a maximum strain of 8 µε only, in 2 µε increments. The recordings of the SAW and reference strain gauge signals are shown in Fig. 5, and follow similar trends. The SAW sensitivity is roughly 0.6 ppm/µε.

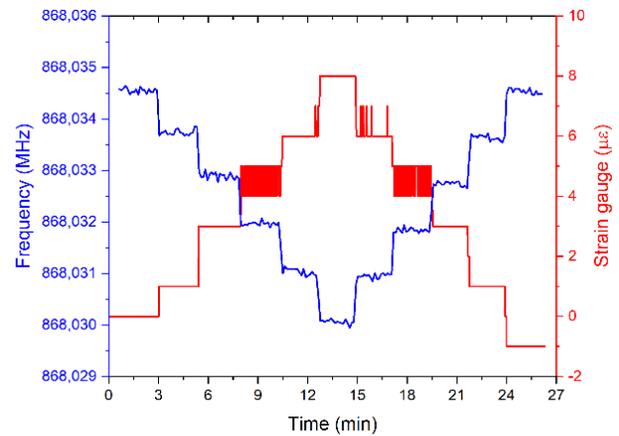

Fig. 5. Wired SAW and reference strain gauge measurement data (rebar 2), during low strain cycle. Blue: resonant frequency of the wired SAW resonator; Red: reference strain gauge readings.

### B. Tests in bending machine

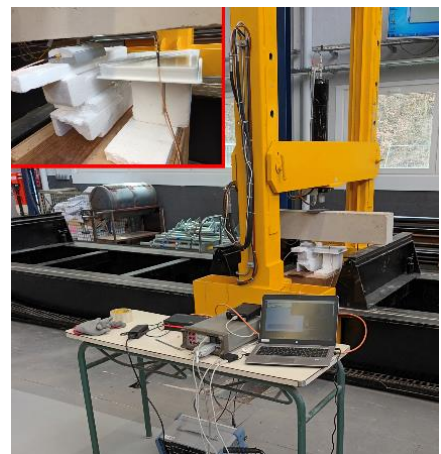

Fig. 6. Measurement setup during bending tests. Inset: mounting configuration of the interrogation antennas.

The next series of tests was conducted at higher loads, using a hydraulic bending machine. The machine can apply a maximum load of 400 kN, with an average loading speed of 0.06 kN/s (see Fig. 6).

The load was applied automatically, and changed every 30 seconds, to follow pre-programmed load cycles. The measured data from the wireless SAW strain sensor and refence strain gauge are shown in Fig.7. The 868 MHz antenna (seen on the right side of Fig. 6's inset) was placed 7 cm away from the surface of the beam. The pre-load applied by the bending machine at the start and at the end of the test can clearly be seen, in the data. The pre-load resulted in a deformation level of up to +500 µε, on the rebar. The maximum strain was ~1050 µε. The SAW signal is noisy. However, it does match that of the reference gauge.

In the last test, a strain of up to 850 µε was applied, in 100 µε increments. The data from the wired SAW (mounted on rebar 2) and reference strain gauge is shown in Fig. 8 (both curves follow, again, similar trends). There is again a good overall correlation between both readings. The higher the strain, the more the SAW signal drifts as a function of time (at constant load). This is likely due to relaxation effects, in the cyanoacrylate glue used to fixate the SAW die on steel.

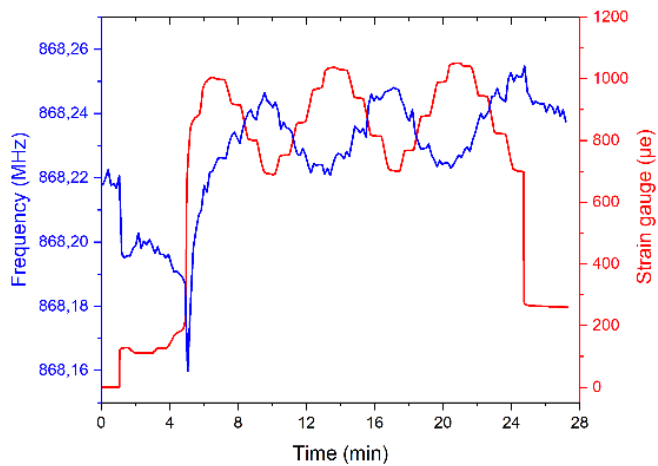

Fig. 7. Wireless SAW and reference strain gauge measurement data (rebar 1), during high strain cycle. Blue: resonant frequency of the wireless SAW resonator; Red: reference strain gauge readings. The reader antenna was located 7 cm under concrete surface.

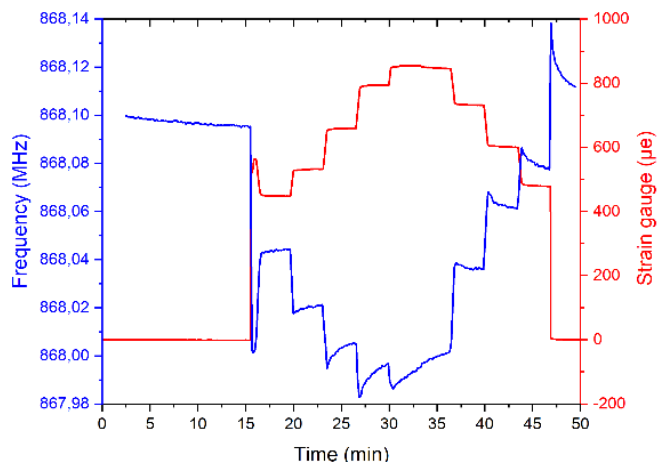

Fig. 8. Wired SAW and reference strain gauge measurement data, during high strain cycle (rebar 2). Blue: resonant frequency of the wired SAW resonator; Red: reference strain gauge readings.

## V. CONCLUSION AND DISCUSSION

We demonstrated that commercial SAW sensors can be used, in wired as well as wireless configurations, to measure temperature and strain inside concrete elements. The accuracy of the measurements improves over time, as concrete hardens. Wired SAW strain sensors showed a very good stability in the low strain regime. However, at high strain, relaxation and creep occur within the glue used to mount the sensors on steel, which result in strong signal drifts (as reported by Kalinin [9]).

In future work, we want to improve the RF link, by using better antennas and improved antenna tuning (by considering the dielectric properties of concrete, the metallic reinforcements and the antennas' protecting epoxy layers). We also want to reduce the drift at high strain, following three possible approaches: 1. the use of a better adhesive, 2. the use of Nanofoil to achieve glue-less bonding [10], and 3. the direct integration of piezoelectric thin films and SAW sensors onto metallic parts [11].

## ACKNOWLEDGMENT

This work was partially supported by Project i-MON (FFG COIN Funding Program, https://projekte.ffg.at/projekt/3984454).

## REFERENCES

[1] L. Xie, X. Zhu, Z. Liu, X. Liu, T. Wang, and J. Xing, "A rebar corrosion sensor embedded in concrete based on surface acoustic wave," *Measurement*, vol. 165, p. 108118, Dec. 2020, doi: 10.1016/j.measurement.2020.108118.
[2] Y. Kim *et al.*, "Practical wireless safety monitoring system of long-span girders subjected to construction loading a building under construction," *Measurement*, vol. 146, pp. 524–536, Nov. 2019, doi: 10.1016/j.measurement.2019.05.110.
[3] S. Taheri, "A review on five key sensors for monitoring of concrete structures," *Construction and Building Materials*, vol. 204, pp. 492–509, Apr. 2019, doi: 10.1016/j.conbuildmat.2019.01.172.
[4] K. Čápová, L. Velebil, and J. Včelák, "Laboratory and In-Situ Testing of Integrated FBG Sensors for SHM for Concrete and Timber Structures," *Sensors*, vol. 20, no. 6, Jan. 2020, doi: 10.3390/s20061661.
[5] E. DiGiampaolo, A. DiCarlofelice, and A. Gregori, "An RFID-Enabled Wireless Strain Gauge Sensor for Static and Dynamic Structural Monitoring," *IEEE Sensors Journal*, vol. 17, no. 2, pp. 286–294, Jan. 2017, doi: 10.1109/JSEN.2016.2631259.
[6] W. C. Wilson, M. D. Rogge, B. H. Fisher, D. C. Malocha, and G. M. Atkinson, "Fastener Failure Detection Using a Surface Acoustic Wave Strain Sensor," *IEEE Sensors Journal*, vol. 12, no. 6, pp. 1993–2000, Jun. 2012, doi: 10.1109/JSEN.2011.2181160.
[7] J. Streque *et al.*, "Design and Characterization of High-Q SAW Resonators Based on the AlN/Sapphire Structure Intended for High-Temperature Wireless Sensor Applications," *IEEE Sensors Journal*, vol. 20, no. 13, pp. 6985–6991, Jul. 2020, doi: 10.1109/JSEN.2020.2978179.
[8] V. Kalinin, A. Leigh, A. Stopps, and S. B. Hanssen, "SAW torque sensor for marine applications," in *2017 Joint Conference of the European Frequency and Time Forum and IEEE International Frequency Control Symposium (EFTF/IFCS)*, Jul. 2017, pp. 347–352. doi: 10.1109/FCS.2017.8088889.
[9] V. Kalinin, "Modelling of hysteresis and creep in SAW strain sensors," in *2014 IEEE International Frequency Control Symposium (FCS)*, May 2014, pp. 1–4. doi: 10.1109/FCS.2014.6859976.
[10] P. Nicolay, J. Bardong, H. Chambon, and P. Dufilie, "Glue-Less and Robust Assembly Method for SAW Strain Sensors," in *2018 IEEE International Ultrasonics Symposium (IUS)*, Oct. 2018, pp. 1–4. doi: 10.1109/ULTSYM.2018.8580224.
[11] P. Mengue *et al.*, "Direct integration of SAW resonators on industrial metal for structural health monitoring applications," *Smart Mater. Struct.*, vol. 30, no. 12, p. 125009, Oct. 2021, doi: 10.1088/1361-665X/ac2ef4.